\documentclass[final,3p,times]{elsarticlemod}

\usepackage{amssymb}
\usepackage{amsfonts}

\usepackage{graphicx}

\usepackage{amsthm}
\newtheorem{thm}{Theorem}

\newtheorem{cor}[thm]{Corollary}

\newdefinition{defn}{Definition}

\begin{document}

\begin{frontmatter}

\title{An Algebra to Merge Heterogeneous Classifiers}

\author[1]{Philippe J. Giabbanelli}\corref{cor1}
\ead{pg438@cam.ac.uk}

\author[2]{Joseph G. Peters}
\ead{peters@cs.sfu.ca}

\address[1]{University of Cambridge, Cambridge, CB2 0QQ, United Kingdom}

\address[2]{School of Computing Science,
Simon Fraser University, Burnaby, British Columbia, V5A 1S6, Canada}

\cortext[cor1]{Corresponding author. {\em Phone:} +44 (0)1223 330315, {\em FAX:} +44 (0)1223 330316}





\begin{abstract}
In distributed classification, each learner observes its environment and deduces a classifier. As a learner has only a local view of its environment, classifiers can be exchanged among the learners and integrated, or \textit{merged}, to improve accuracy. However, the operation of merging is not defined for most classifiers. Furthermore, the classifiers that have to be merged may be of different types in settings such as ad-hoc networks in which several generations of sensors may be creating classifiers. We introduce {\em decision spaces} as a framework for merging possibly different classifiers. We formally study the merging operation as an algebra, and prove that it satisfies a desirable set of properties. The impact of time is discussed for the two main data mining settings. Firstly, decision spaces can naturally be used with non-stationary distributions, such as the data collected by sensor networks, as the impact of a model decays over time. Secondly, we introduce an approach for stationary distributions, such as homogeneous databases partitioned over different learners, which ensures that all models have the same impact. We also present a method that uses storage flexibly to achieve different types of decay for non-stationary distributions. Finally, we show that the algebraic approach developed for merging can also be used to analyze the behaviour of other operators.
\end{abstract}

\begin{keyword}
Model combination; Non-stationary distributions; Unsupervised meta-learning




\end{keyword}

\end{frontmatter}


\section{Introduction}
\label{intro}

A wide variety of systems, such as sensor and peer-to-peer networks, are composed of independent entities. These entities record observations from their local environments, for example as databases of observed tuples. The entities often have to adapt to future changes, but their local views may be too restricted to compute classifiers that can make accurate predictions. Therefore, a global classifier that can be used to predict global trends is often desirable. Entities might not compute a global classifier by exchanging their datasets directly, because of security concerns, or because the large quantity of information could overwhelm limited resources such as local storage or battery power. Thus, the global classifier has to be realized by combining the classifiers deduced by the entities. The need to combine classifiers also arises in other situations. Several examples illustrate the need to combine classifiers in Section~\ref{examples}, and others can be found in the series of annual International Workshops on Multiple Classifier Systems~\cite{SKR11}.

Different ways have been proposed to realize a global classifier (e.g., ensemble classifiers, meta-learning). These ways have been grouped differently, but a key distinction can be found between \textit{fusion} and \textit{selection} (~\cite{Kun04}, p. 106). Intuitively, classifiers are specialized when they are restricted to instances with specific features: when a new instance comes in, the appropriate classifier is thus selected. Alternatively, classifiers may be designed for instances with the same features, and they are then fused. In this paper, we are concerned with the intermediate case: classifiers can have an overlapping feature space, but some may be more competent in parts of that space than others. This is typically reflected in approaches that weight the outputs of the classifiers depending on the instance being classified~\cite{Kun04}. To emphasize that our approach is intermediate, we say that we \textit{merge} classifiers instead of selecting or fusing them.

Merging classifiers is a complicated operation for several reasons. Firstly, classifiers can be of different structures, such as decision trees and support vector machines. Secondly, conflicts arise when classifiers differ on their predictions. Solving these conflicts either uses heuristics (e.g., pure meta-learning~\cite{CS93,VD02,CVB04}), or requires access to the dataset used to train the classifiers, which is undesirable in settings such as sensor networks due to the high cost of transmission. In this paper, we consider the case of meta-learning in which the classifiers are potentially of different types. The concept of meta-learning and the precise types are described in Section~\ref{background}.

Most of the previously proposed solutions have been application-specific and were validated through experiments, leading researchers to suggest that questions about combining classifiers are (still) open ``because we do not yet have a scientific understanding of the classifier combination mechanisms''~\cite{Ho02}. Consequently, our approach emphasizes the formal aspects. In Section~\ref{decisionspaces}, we introduce decision spaces as an algebraic framework to investigate the behaviour of distributed data mining algorithms in which learners propagate only local models and not local observations. This framework can be used in a large variety of cases since it allows the combination of different models ({\em e.g.,} decision trees, rule sets, support vector machines with linear kernels) and does not rely on homogeneous observations. In Section~\ref{merge}, we present a heuristic merge operator that solves all conflicts between models without using their observations. We use the algebra to prove that the operator satisfies a set of desirable properties, such as commutativity: when two models are merged, the result does not depend on which model comes first. However, the operator is not associative: if three or more models have to be merged, then the order in which the operator is applied affects the result. This can be a desirable behaviour in settings such as \textit{data streams}~\cite{MUT05}, in which the distribution changes over time and the most recently received data is considered to be the most representative of current trends.

In other situations, such as homogeneous distributions, this behaviour is not appropriate. For example, a massive homogeneous database could be partitioned among computation units for the sake of efficiency, and the contribution of each unit to the final result should not depend on the time at which it sends its model. In Section~\ref{time}, we develop two protocols, or \textit{schemes}. The first one uses storage flexibly to achieve several types of decay in a data stream setting, and can be used for the case of a homogeneous database. The second scheme uses no storage space and has no decay.

Our formal framework can also be used to characterize other common, yet difficult, problems of data mining. For example, a learner can generate a sequence of models and analyze this sequence to find patterns of changes in the underlying system. In a data stream setting, this is referred to as a \textit{blind method} operating over a sliding window. In Section~\ref{algebra}, we define a reduction operator that reduces a sequence of models to a single model to permit easier analysis of the sequence, and we characterize its properties using an algebraic approach. We also briefly discuss how more complex operators can be defined using compositions of the reduction and merge operators.

\section{Applications}
\label{examples}
\subsection{Combining maps}
Let us consider that we are monitoring the incidence of a chronic disease in a city. For each part of the city, we want to know whether the incidence is high enough to prompt a specific action. Different groups provide maps of the disease, but these groups might not use the same spatial unit. One group may have divided the city into blocks while another one uses administrative borders. Furthermore, reports may have been released in different months, and administrative borders may have evolved. Thus, we need to merge maps reporting on the incidence with different units. Specifically, we do not have access to individual data (i.e., the data points upon which the maps are based), and we must guarantee that all maps have the same impact in our composite picture. The decision space framework that we propose is able to handle these requirements,
and the algebra can be used to prove properties of the composite map.
Our framework can also handle the situation in which the monitoring would be done over a long period of time, and more recent maps need to have a higher impact in the composite map. Note that, in our framework, each element of the map has to be a polygon; if a region is delimited by curves then it will have to be approximated.

\subsection{Map algebras}
The problem of combining maps is classically faced by geographical information systems (GIS). In the early 1980s, Tomlin proposed to consider maps as two-dimensional arrays and to design script-like languages to manipulate them in GIS~\cite{Tom83}. These languages allow operations such as the selection of areas (e.g., $\mathit{School}_{\mathit{area}}=\mathit{distance(SCHOOLS)}<300m$), as well as the combination of selected areas using numerical operations. Thus, map algebras are specialized programming languages for manipulating the cells of grids, which now offer the possibility to process maps within GIS using cellular automata or image-processing techniques~\cite{CCF09,Pul01}. While these algebras can be used to combine maps~\cite{Fra05}, there is no (mathematically proven) control of the impact that each map has on the final combination.

\subsection{Wireless sensor networks}
Wireless sensor networks~\cite{ASYC02} are a natural application for our approach. They can be found in applications such as forest fire detection, where temperature sensors measure the current temperature and its rate of change. They are also used to study earthquake activity by measuring the strength and duration of seismic waves in the earth's crust. The measurements are forwarded to collection nodes, which forward the data to a base station for processing. A typical sensor has a modest battery life which can be quickly drained by the transceiver when sending large amounts of data. Thus it is infeasible to forward all of the collected data to the base station. However, sensors have the computing capabilities to perform summaries such as averages and standard deviations which can then be sent to the base station.

In our approach, each sensor can construct a classifier based on its local observations, and send it as a summary. This summary is richer than a simple average when the goal is to make predictions. This richness may come at a cost in some classifiers: for example, a model derived using random forest classifiers~\cite{B01} may be heavier than its training data, and would not be appropriate for wireless sensor networks. However, the model is usually lighter when it simplifies the data (e.g., using pruning in decision trees), and we focus on classifiers that lead to such models. Once a classifier is sent by each sensor, a collection node can merge the classifiers that it receives to form a model of a region before forwarding it toward the base station. The merging of classifiers provides a trade-off between battery consumption and prediction accuracy. The merging improves the accuracy of predictions compared to sending simple averages in exchange for an increase in battery consumption. The merging also provides a transparent means to give greater weight to the most recent observations when the network is deployed over a long time.

\section{Background}
\label{background}

Our work focuses on classifiers. A classifier learns from a dataset of example instances about how target attributes are based on the values of predictive attributes. Then, it can address the classification problem of predicting the target attributes of a new instance based on the values of its predictive attributes. Without loss of generality, we explain our framework by focussing on one target attribute, 
but the operations can be applied for any number of target attributes.

Formally, an instance has the form
$(a_1, \ldots, a_m, y)$ where $a_i$ is the value of the $i$-\textit{th} attribute, and $y$ is a class label for this data. For example, suppose that there are two instances $(a_1 = 30, y = \mathit{No})$ and $(a_1 = 70, y = \mathit{Yes})$ where attribute $a_1$ is an age in $[0, 125]$ and the class label is either $\mathit{Yes}$ or $\mathit{No}$, corresponding to an individual being classified as old or not. The goal is to find the best approximation to the function $f(a_1, \ldots, a_m) = y$ that determines the label of an unclassified instance given the values of its attributes. Three of the most commonly used techniques to train a classifier, \textit{i.e.,} to deduce an approximation of $f$ given a set of instances, are~\cite{HK06,HAM09}
\begin{itemize}
	\item A \textit{Support Vector Machine} (SVM) classifies the instances into two classes by separating them with the $(m - 1)$-dimensional hyperplane that leaves the maximum margin between the two classes~\cite{SLS99}. It is also possible to obtain non-linear classifications using the kernel method~\cite{BGV92}.
	\item One of the most popular classifiers is the {\em decision tree}.
A {\em decision tree learner}~\cite{QUI96} applies a divide-and-conquer technique to recursively split the data using the value of an attribute while maximizing a metric such as the information gain or Gini index. Each split is represented as a node and the recursive procedure yields a tree. Thus, a path in the tree corresponds to a set of conditions on the values of the attributes, and leads to a class distribution vector expressing the percentage of instances in each class.
	\item A classifier can also be a set of rules based on the values of the attributes. This is commonly referred to as a \textit{rule set}~\cite{FAR05}. A rule is a conjunction of conditions on the attributes that results in a class distribution vector as in a decision tree. An attribute can be repeated at most twice in a rule, to specify a lower bound and an upper bound of an interval. 
\end{itemize}

In meta-learning, each learner builds a classifier using only its own data. For example, each unit in a network of radar sensors can collect data about the movements of targets within its radius. Then, each sensor uses its own data to deduce a classifier such as a decision tree. In such an environment, all sensors may want to have a classifier that is as accurate as possible and thus they exchange their classifiers and merge them.
In another setting, one may want to achieve a speedup by using parallel algorithms and thus a homogeneous database is partitioned among different computation units, either by providing them with subsets of observations or subsets of attributes (vertical partitioning~\cite{SM08}). Then, the units send their classifiers to common sources in charge of merging.
In both homogeneous databases and distributed environments, the problem is to merge a set of classifiers to create a single classifier.

A number of researchers have focussed on merging decision trees: it was noted in~\cite{BKS05} that {\em ``a kind of decision tree induction [that is] efficient in a wide area system employs meta-learning, [in which] each computer induces a decision tree based on its local data and then the different models are merged to form the final tree"}. For example, it has been proposed~\cite{HCB98} to transform decision trees into the sets of rules that they represent and to merge those sets. A rule not in conflict with other rules is kept intact and otherwise the conflicts are resolved using a heuristic. However, there are two potential problems with the design of the heuristic proposed in~\cite{HCB98}. Firstly, the authors argue that conflicts that are not handled by the heuristic are {\em ``unlikely if the training sets contain similar distributions of examples from a coherent larger training set"}, thus the approach is limited to homogeneous data bases. Secondly, their algorithm could ask a learner to send all of the data on which there is a conflict in order to perform data mining again, and this prevents uses in settings such as sensor networks.

The requirement that the data needs to be examined in case of conflicts is found in several other approaches. For example, arbiter meta-learning~\cite{TPT99} is a technique that merges classifiers in a hierarchical way. This technique has been designed for homogeneous databases, and parts of the data must be propagated during the combination process. Ganti and colleagues~\cite{GGR02} focused on a different problem: they examined how one could quantify the difference between two datasets by analyzing their models, which can be used to determine whether a model should be updated. Their concepts are similar to ours in that they considered a model as a geometrical space divided into units, each of which is assigned a value, and combining two models requires scanning the data. The key difference of our framework, introduced in the next section, is its ability to merge potentially heterogeneous classifiers without requiring any data besides the models.

\section{A framework: decision spaces}
\label{decisionspaces}
\subsection{Introducing the structure}
Intuitively, a \textit{decision space} is an $m$-dimensional space, in which each dimension corresponds to one of the $m$ attributes. It contains a set of non-overlapping elements which, if they cover all the space, form a partition of the space. A geometrical interpretation of an element is a subspace defined by an $m$-\textit{polytope}. Thus, the only requirement for the classifiers considered in this paper is that their elements have to be polytopes, which is the case for SVMs with linear kernels, decision trees, and rule sets. Our study of polytopes is a first step to investigating the theoretical behaviour of pure meta-learning. Further research could remove the restriction to polytopes to generalize our approach to handle other classifiers such as SVMs with non-linear kernels~\cite{BGV92}. Informally, non-linear kernels can define shapes in terms of curves whereas polytopes can only use lines.

\begin{figure}[t]
\centering
\includegraphics[scale=.35]{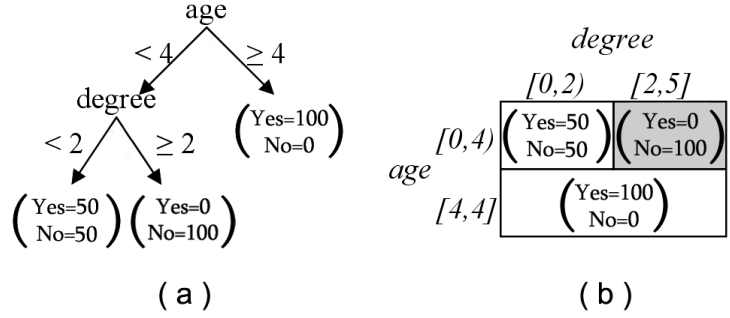}
\caption{Decision tree (a) and decision space (b) representations.}
\label{4-Comparison}
\end{figure}

Each element, or polytope, has a class distribution vector that specifies the percentage of instances within the covered space that are assigned to each class. Determining the percentages by enumerating the instances is straightforward for the three classifiers considered here. An example of a decision space is shown in Figure~\ref{4-Comparison}(a): it has two attributes, $degree$ and $age$, and three elements, each with a class distribution vector of size 2 (with classes $\mathit{Yes}$ and $\mathit{No}$). Any classifier is capable of producing a label for a given instance, and can be repeteadly prompted for labels over a discrete space. Therefore, any classifier can provide a class distribution vector. In Xu's categorization of the outputs used to combine classifiers, our requirement is the most universal (Type 1)~\cite{XKS92}. These concepts are formalized in Definitions~\ref{Def1} and~\ref{Def2}.

\begin{defn}
{\rm A decision space is an $m$-dimensional (product) space $D_{attr_1} \times \cdots \times D_{attr_i} \times \cdots \times D_{attr_m}$ where $m$ is the number of attributes and $D_{attr_i}$ is the space covered by the $i$-\textit{th} attribute specified as a bounded poset ({\it i.e.,} a partially ordered set with a least and a greatest element).}
\label{Def1}
\end{defn}

\begin{defn}
{\rm An element of an $m$-dimensional decision space $D$ is a subspace of $D$, \textit{i.e.,} an $m$-dimensional polytope ({\em m-polytope}). It is identified by a set of coordinates for each attribute, and has a class distribution vector $V$ with $c$ components,
where $c$ is the number of classes. The $i$-{\em th} component of the vector is the percentage of instances in class $i$, which is obtained by counting all instances that have class $i$ and are within the element's space. Thus, the content of the vector sums to 100\%.} We will refer to the vector $V$ as the {\em value} of the element.
\label{Def2}
\end{defn}

\subsection{Conversions}
The constraints on the structure of a classifier are used by a data mining algorithm to guide the search. A decision space is a framework and does not result directly from a data mining algorithm, thus its structure is less constrained than classifiers such as decision trees. This ensures that elements from several kinds of classifiers can be converted to elements of a decision space with no loss of information. To convert a classifier into a decision space, the only data required besides the classifier itself are the ranges of attributes for the dataset on which the classifier was trained. These ranges can be deduced in one pass over the dataset by scanning for the maximum and minimum values. The ranges can also be user-supplied, but should not be smaller than what is found in the dataset for consistency sake. Thus, given a classifier and the ranges of the attributes, the main task of the conversion is to extract the individual elements, or polytopes. The polytopes for elements of the three classifiers defined in Section~\ref{background}, in order of increasing constraints on the shapes, are as follows:
\begin{itemize}
	\item[(1)] The elements in an SVM can have the most general shapes because the data can be assigned into regions.
	\item[(2)] A rule in a rule set defines an axis-parallel rectangle. 
	\item[(3)] Each path of a decision tree (Figure~\ref{4-Comparison}(b)) can be converted to a rule, and  this rule defines an axis-parallel rectangle (Figure~\ref{4-Comparison}(a)).
\end{itemize}

A decision tree cannot generate all axis-parallel rectangles. Indeed, a decision tree belongs to the data mining family of \textit{divide-and-conquer} algorithms that imposes constraints on the search. Intuitively, a cut in the space along the border of an element, either vertical or horizontal, should not cut any element~\cite{FUR99}. An example of a set of rules that violates this constraint is given below, and shown in Figure~\ref{4-Partition}.
\begin{center}{\sc IF} $age \geq 0$ {\sc and} $age < 4$ {\sc and} $degree \geq 0$ {\sc and} $degree < 2$ {\sc THEN} $A$\\
{\sc IF} $age \geq 4$ {\sc and} $age < 6$ {\sc and} $degree \geq 0$ {\sc and} $degree < 4$ {\sc THEN} $B$\\
{\sc IF} $age \geq 0$ {\sc and} $age < 2$ {\sc and} $degree \geq 2$ {\sc and} $degree < 6$ {\sc THEN} $C$\\
{\sc IF} $age \geq 2$ {\sc and} $age < 6$ {\sc and} $degree \geq 4$ {\sc and} $degree < 6$ {\sc THEN} $D$\\
{\sc IF} $age \geq 2$ {\sc and} $age < 4$ {\sc and} $degree \geq 2$ {\sc and} $degree < 4$ {\sc THEN} $E$
\end{center}

\begin{figure}[t]
\centering
\includegraphics[scale=.35]{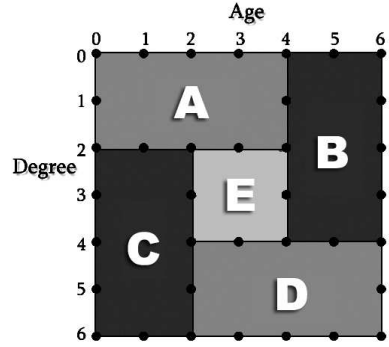}
\caption{A partition of space not allowed by decision trees but allowed by decision spaces.}
\label{4-Partition}
\end{figure}

The elements of rule sets and decision trees can be converted to elements of a decision space using Algorithm~1. 
The algorithm uses pattern-matching. For example, in line 4, $attr_k$ $OP_1=\{<,\leq\}$ $val_{k_{up}}$ is a pattern for which the value of an attribute has to be lower or strictly lower than the value $val_{k_{up}}$. If the pattern is found, then $attr_k$, $OP_1$, and $val_{k_{up}}$ are bound to the actual values. For example, a pattern $age < 5$ will result in the binding $attr_k = age$, $OP_1 = ``<"$, $val_{k_{up}} = 5$.

For each rule, the algorithm considers all patterns that specify an upper bound (line 4). If there is also a pattern specifying a lower bound for the same attribute (line 6), then the range of the attribute can be specified. If no pattern is found for the lower bound of an attribute $attr_k$ (line 15), then we use the lower bound of the attribute's range which we denote $min(D_{attr_{k}})$. Finally, if no upper bound is found for $attr_k$ (line 20), we use the upper bound of the attribute's range which we denote $max(D_{attr_{k}})$.

Converting elements from a support vector machine follows a simpler process: each element of the decision space corresponds to exactly one element of the SVM with the same distribution vector and the same coordinates specifying the space spanned.\\

\noindent
\begin{minipage}{\textwidth}
\label{Alg1}
\newlength{\algcoma}
\setlength{\algcoma}{7.5cm}
\hrulefill\\
{\bf Algorithm 1}\quad RulesToSpaces(Ruleset $R$, Attribute ranges $D_{attr1} \times \cdots \times D_{attrM}$)\\

\vspace*{-6mm}\hrulefill\\
{\bf Require:} Rules are expressed in the following form:\\
\hspace*{7mm}$r := $ {\sc IF} $attr_1 \asymp val_{1}$ {\sc AND $\cdots$ AND} $attr_m \asymp val_{m}$ {\sc THEN} $class = X$ \\
\vspace*{-5mm}\begin{tabbing}
xxx\=xx\=xx\=xx\=xx\=XXXXXXXXXXXXXXXXXXXXXXXXXX\= \kill
\ \,1.\>Decision space $S \leftarrow \emptyset$\>\>\>\>\`// \parbox[t]{\algcoma}{Initially, the decision space is empty}\\
\ \,2.\>{\bf for} $r \in R$ {\bf do}\>\>\>\>\`// \parbox[t]{\algcoma}{Each rule of the rule set is converted to one element}\\
\ \,3.\>\>$element$ $e.value \leftarrow r.X$\>\>\>\`// \parbox[t]{\algcoma}{The element’s value is the one predicted by the rule}\\
\ \,4.\>\>$P \leftarrow$ {\sc all patterns} $attr_k$ $OP_1=\{<,\leq\}$ $val_{k_{up}}$ {\sc in} $R$\>\>\>\`// \parbox[t]{\algcoma}{A rule is a conjunction of patterns, each delimiting a bound}\\
\ \,5.\>\>{\bf for} $p \in P$ {\bf do}\>\>\>\`// \parbox[t]{\algcoma}{Use each upper bound to find the element’s space}\\
\ \,6.\>\>\>{\bf if} {\sc $\exists$ a pattern} $attr_k$ $OP_2=\{>,\geq\}$ $val_{k_{low}}$ {\sc in} $R$ {\bf then}\>\>\`// \parbox[t]{\algcoma}{Check if there is a lower bound on the same variable}\\
\ \,7.\>\>\>\>{\bf if} $OP_1$ {\sc is} $<$ {\sc and} $OP_2$ {\sc is} $>$ {\bf then}\>\`// \parbox[t]{\algcoma}{The two bounds do not include the endpoints}\\
\ \,8.\>\>\>\>\>$e.space \leftarrow e.space \cup (val_{k_{low}}, val_{k_{up}})$\\
\ \,9.\>\>\>\>{\bf else if} $OP_1$ {\sc is} $<$ {\sc and} $OP_2$ {\sc is} $\geq$ {\bf then}\>\`// \parbox[t]{\algcoma}{Only the upper endpoint of the bound is included}\\
10.\>\>\>\>\>$e.space \leftarrow e.space \cup (val_{k_{low}}, val_{k_{up}}]$\\
11.\>\>\>\>{\bf else if} $OP_1$ {\sc is} $\leq$ {\sc and} $OP_2$ {\sc is} $>$ {\bf then}\>\`// \parbox[t]{\algcoma}{Only the lower endpoint of the bound is included}\\
12.\>\>\>\>\>$e.space \leftarrow e.space \cup [val_{k_{low}}, val_{k_{up}})$\\
13.\>\>\>\>{\bf else}\>\`// \parbox[t]{\algcoma}{Both endpoints of the bound are included}\\
14.\>\>\>\>\>$e.space \leftarrow e.space \cup [val_{k_{low}}, val_{k_{up}}]$\\
15.\>\>\>{\bf else}\>\>\`// \parbox[t]{\algcoma}{We only have an upper bound on the element}\\
16.\>\>\>\>{\bf if} $OP_1$ {\sc is} $<$ {\bf then}\\
17.\>\>\>\>\>$e.space \leftarrow e.space \cup [min(D_{{attr}_k}),val_{k_{up}})$\\
18.\>\>\>\>{\bf else}\\
19.\>\>\>\>\>$e.space \leftarrow e.space \cup [min(D_{{attr}_k}),val_{k_{up}}]$\\
20.\>\>{\bf if} $P = \emptyset$ {\bf then}\>\>\>\`// \parbox[t]{\algcoma}{If there was no upper bound}\\
21.\>\>\>$P \leftarrow$ {\sc all patterns} $attr_k$ $OP=\{>,\geq\}$ $val_{k_{low}}$ {\sc in} $R$\>\>\`// \parbox[t]{\algcoma}{Then access the lower bound}\\
22.\>\>\>\>{\bf for} $p \in P$ {\bf do}\\
23.\>\>\>\>{\bf if} $OP_2$ {\sc is} $>$ {\bf then}\>\`// \parbox[t]{\algcoma}{The lower bound is excluded}\\
24.\>\>\>\>\>$e.space \leftarrow e.space \cup (val_{k_{low}}, max(D_{{attr}_k})]$\\
25.\>\>\>\>{\bf else}\\
26.\>\>\>\>\>$e.space \leftarrow e.space \cup [val_{k_{low}}, max(D_{{attr}_k})]$\`// \parbox[t]{\algcoma}{The lower bound is included}\\
27.\>\>$S \leftarrow S \cup e$\>\>\>\`// \parbox[t]{\algcoma}{Add the element to the decision space}\\
\end{tabbing}
\end{minipage}

\section{Merge operator}
\label{merge}

\subsection{Preliminary definitions}

The most fundamental operation on decision spaces is merge. In the following, we use the term ``element" to refer to a geometrical space spanned, and ``value" to refer to a class distribution vector. Given two decision spaces $X$ and $Y$, we merge them into $Z$ using the following principles: \\

\noindent
\fbox{
\begin{minipage}[t]{.974\textwidth}
{\bf Merge principles}
\begin{enumerate}
	\item[(1)] If an element (subspace) $x \in X$ does not intersect with any element $y \in Y$, then the prediction represented by $x$ has no conflicts and can be added to $Z$. An element $y \in Y$ with no conflicts is treated similarly.
	\item[(2)] If an element $y \in Y$ is strictly contained within an element $x \in X$ \textit{with the same value} (\textit{i.e.,} the same class distribution vector $V$), then we consider $y$ to be too specialized (as explained in Definition~\ref{Def5}) and delete it.
	\item[(3)] If neither of the first two conditions is satisfied, then the element $x \in X$ intersects with at least one element of $Y$ and conflicts must be resolved.
\end{enumerate}
\end{minipage}
} \\

Prior to establishing the algorithm to merge elements, we introduce the formal notation on which it relies, and specify the ways that elements can intersect. From here on, we use the following notation for an element $x \in X$:
\begin{itemize}
	\item $x$ has a set of attributes $A(x)$.
	\item Each attribute $a \in A(x)$ covers a (one-dimensional) space $S(x,a)$. It can be an interval such as $[8, 10]$, or a union of intervals.
	\item The size of the space covered by $x$ for the attribute $a$ is denoted $|S(x, a)|$.
	\item $x$ has a class distribution vector $V(x)$.
\end{itemize}

\begin{defn}
{\rm An element $x \in X$ {\em subsumes} an element $y \in Y$, denoted $y \preceq x$, if $A(x) = A(y)$ and $\forall a \in A(x)$, $S(y, a) \subseteq S(x, a)$.}
\label{Def3}
\end{defn}

In other words, an element $y$ is subsumed by an element $x$ when the space covered by $y$ is included in the space covered by $x$. We denote strict subsumption by $\prec$ when $\forall a \in A(x)$, $S(y,a) \subset S(x,a)$. The main property of subsumption is established by Theorem~\ref{Th1}.

\begin{thm}
Let $X$ and $Y$ be two decision spaces. For each $y \in Y$, there is at most one $x \in X$ such that $y \preceq x$.
\label{Th1}
\end{thm}

\begin{proof} The elements of $X$ and $Y$ partition the spaces formed by $X$ and $Y$, respectively. Thus, one element (\textit{i.e.,} component of a partition) can be completely included in at most one other element.\label{Proof1}\end{proof}

\begin{defn}
{\rm The {\em intersection} of an element $x \in X$ with a decision space $Y$ is denoted $x \uplus Y$ and is the largest subset of elements $I = \{y_1, \ldots, y_n\} \subseteq Y$ such that for each $y_i \in I$, $\exists a \in (A(x) \cap A(y_i))$ such that $S(x,a) \cap S(y_i,a) \neq \emptyset$. Subsumption is a special case of intersection.}
\label{Def4}
\end{defn}

The first two principles of merging can be handled by the notions in Definitions~\ref{Def3} and~\ref{Def4}. For the third principle, we resolve each conflict between two elements $x \in X$ and $y \in Y$ by creating a new element $z$ for each intersection. A simple approach would be to assign to $z$ a value that is the average of the values of $x$ and $y$, but it would not take into consideration the spaces covered by $x$ and $y$ (\textit{i.e.}, their regions of competence). Consequently, Definition~\ref{Def5} specifies how to measure the space covered by an element, which we call its \textit{specialization} (also known as competence). Then, Definition~\ref{DefMetric} calculates the value of $z$ is calculated as a weighted average based on the spaces covered by $x$ and $y$. It should be noted that Definition~\ref{DefMetric} is a classical combination scheme intermediate between classifier fusion and selection (\textit{cf}.\ Introduction). In these schemes, all classifiers contribute to the outcome of a given space space with weights based on their competence for that space~\cite{Kun04}.

\begin{defn}
The {\em specialization} of an element $x \in X$ is
${\displaystyle M(x) = \frac{\sum{_{a \in A(X)} |S(x,a)|}}{|A(X)|}.}$
\label{Def5}
\end{defn}

Intuitively, we sum the sizes of the spaces of the attributes characterizing $x$ and we normalize by the number of attributes. Small values of $M$ indicate specialized predictions based on small spaces. Other possible metrics could take into account the number of instances, or the class distribution vector. However, in the case of pure meta-learning investigated in this paper, we cannot use the number of instances or the class distribution vector in a specific part of an element. This could only be done by requiring the original classifier to perform an exhaustive search in its dataset, and repeated use of such an operation could lead to significant overhead. Therefore, such alternative metrics are not considered here.

\begin{defn}
The {\em value} of the intersection of two elements $x \in X$ and $y \in Y$ based on their specializations is the class distribution vector
\[V(x \otimes y) = V(x) \times \frac{M(x)}{M(x)+M(y)} + V(y) \times \frac{M(y)}{M(x)+M(y)}.\]
\label{DefMetric}
\end{defn}
	
If the elements predict several attributes, then the class distribution vector for each attribute is computed using the above formula.
Other formulae for $V(x \otimes y)$ could be designed to suit application-specific needs. Indeed, both the element's space and its value must be taken into account, but specific applications may require that the specialization be normalized differently. However, care should be taken to ensure that custom formulae satisfy the commutativity, idempotency, and unique identity properties. Indeed, such properties are critical for proving that the merge algorithm behaves appropriately, as will be shown in Section~\ref{algprop}.

\subsection{Merging algorithm}
\label{merge2}
\begin{figure}[t]
\centering
\includegraphics[scale=.35]{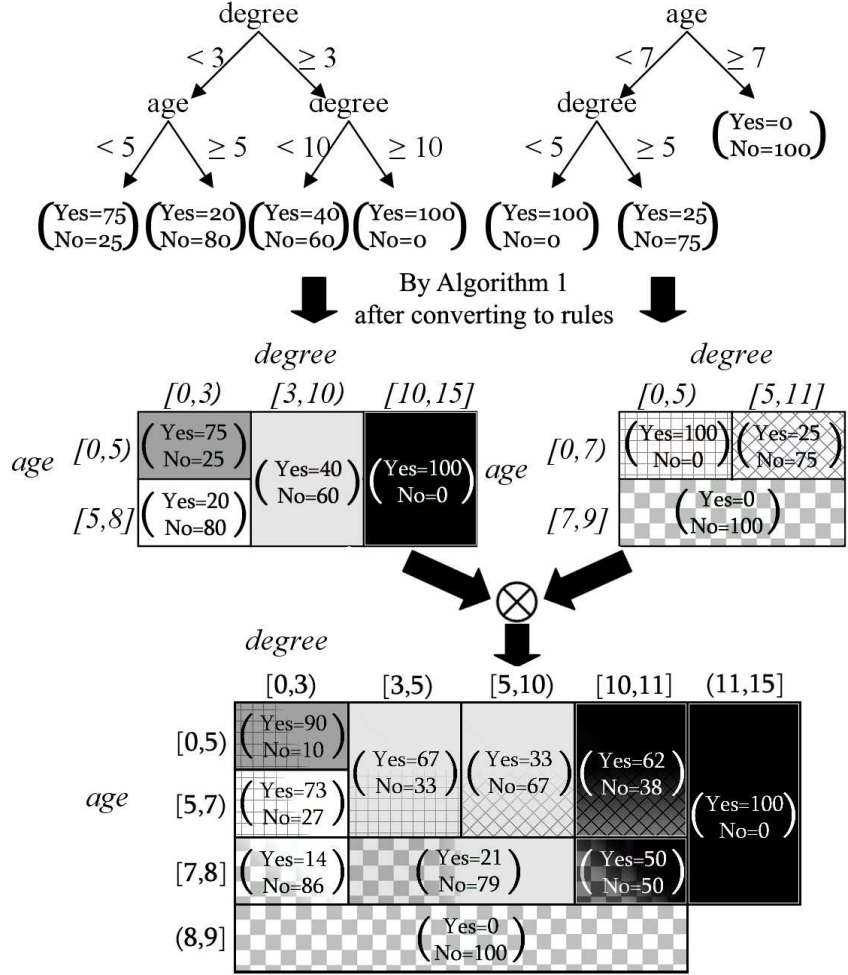}
\caption{Merging two decision trees by converting them into decision spaces and creating a union decision space.}
\label{4-example}
\end{figure}

The merge operator $\otimes: (X, Y) \mapsto Z$ is defined in an algorithmic way by Algorithm~2 
 and is illustrated in Figure~\ref{4-example}. First, we apply Merge principle (1): all elements with no intersection are added. Then, we apply principle (2): each element $x \in X$ that is strictly contained in an element $y \in Y$ with the same value is deleted. Finally, principle (3) is applied as follows:\\

\noindent
\begin{minipage}{\textwidth}
\label{Alg2}
\newlength{\algcomb}
\setlength{\algcomb}{8.4cm}
\hrulefill\\
{\bf Algorithm 2}\quad $\otimes: (X, Y) \mapsto Z$\\

\vspace*{-6mm}\hrulefill\\
\vspace*{-5mm}\begin{tabbing}
xxx\=xx\=xx\=xx\=xx\=xx\=XXXXXXXXXXXXXXXXXXXXXXX\= \kill
\ \,1.\>$Z \leftarrow$ new decision space\\
\ \,2.\>$H \leftarrow \emptyset$\>\>\>\>\>\`// \parbox[t]{\algcomb}{Hash map: set $(y, y')$ of keys $y$ and associated values $y'$}\\
\ \,3.\>{\bf for} $x \in X$ {\bf do}\>\>\>\>\>\`// \parbox[t]{\algcomb}{For each element $x$}\\
\ \,4.\>\>{\bf if} $\nexists y \in Y$ such that $x \prec y$ and $V(x) = V(y)$ {\bf then}\>\>\>\>\`// \parbox[t]{\algcomb}{If there is no $y$ that subsumes it with the same value, then process it}\\
\ \,5.\>\>\>{\bf if} $x \uplus Y = \emptyset$ {\bf then}\>\>\>\`// \parbox[t]{\algcomb}{If $x$ does not conflict with any element of $Y$, then add it to the result}\\
\ \,6.\>\>\>\>$Z \leftarrow Z \cup x$\\
\ \,7.\>\>\>{\bf else}\>\>\>\`// \parbox[t]{\algcomb}{Otherwise, resolve the conflicts}\\
\ \,8.\>\>\>\>$tmp \leftarrow x.space$\>\>\`// \parbox[t]{\algcomb}{The initial space of $x$ is saved}\\
\ \,9.\>\>\>\>{\bf for}	$y \in x \uplus Y$ {\bf do}\>\>\`// \parbox[t]{\algcomb}{Create an element $z$ to handle each conflict}\\
10.\>\>\>\>\>$z \leftarrow$ new element\\
11.\>\>\>\>\>$z.space \leftarrow tmp \cap y.space$\>\`// \parbox[t]{\algcomb}{$z$’s space is the one initially shared by $x$ and the conflicting element}\\
12.\>\>\>\>\>$z.value \leftarrow V(x \otimes y)$\>\`// \parbox[t]{\algcomb}{$z$’s value (class distribution vector) is based on both $x$’s and $y$’s}\\
13.\>\>\>\>\>$Z \leftarrow Z \cup z$\>\`// \parbox[t]{\algcomb}{$z$ is added to the resulting decision space}\\
14.\>\>\>\>\>{\bf if} $\nexists (y,y') \in H$ {\bf then}\>\`// \parbox[t]{\algcomb}{If we never registered a conflict with y then register it}\\
15.\>\>\>\>\>\>$H \leftarrow H \cup (y,y.space \setminus z.space)$\`// \parbox[t]{\algcomb}{Register that the space of $y$ that conflicted has been handled}\\
16.\>\>\>\>\>{\bf else}\>\`// \parbox[t]{\algcomb}{Otherwise update the previous record}\\
17.\>\>\>\>\>\>$H \leftarrow H \setminus (y,y')$\\
18.\>\>\>\>\>\>$H \leftarrow H \cup (y,y'.space \setminus z.space)$\\
19.\>\>\>\>\>$x.space \leftarrow x.space \setminus z.space$\>\`// \parbox[t]{\algcomb}{Also update that some of $x$’s conflicting space was handled}\\
20.\>\>\>\>{\bf if} $x.space \neq \emptyset$ {\bf then}\>\>\`// \parbox[t]{\algcomb}{If, after solving the conflicts, $x$ has (non conflicting) space then we keep it}\\
21.\>\>\>\>\>$Z \leftarrow Z \cup x$\\
22.\>{\bf for} $y \in Y$ such that $\nexists (y,y') \in H$ {\bf do}\>\>\>\>\>\`// \parbox[t]{\algcomb}{If some elements $y$ have never found to be in a conflict $\ldots$}\\
23.\>\>{\bf if} $\nexists x \in X$ such that $y \prec x$ and $V(y) = V(x)$ {\bf then}\>\>\>\>\`// \parbox[t]{\algcomb}{$\ldots$ and that they are not subsumed by an element $x$ with the same value}\\
24.\>\>\>$Z \leftarrow Z \cup y$\>\>\>\`// \parbox[t]{\algcomb}{Then we add them to the result}\\
25.\>{\bf for} $(y,y') \in H$ {\bf do}\>\>\>\>\>\`// \parbox[t]{\algcomb}{For each element $y$ that was found to be in a conflict}\\
26.\>\>{\bf if} $y' \neq \emptyset$ {\bf then}\>\>\>\>\`// \parbox[t]{\algcomb}{If some of this element was free of conflict}\\
27.\>\>\>$Z \leftarrow Z \cup y'$\>\>\>\`// \parbox[t]{\algcomb}{Then we add it}\\
28.\>{\bf return} $Z$
\end{tabbing}
\end{minipage}\\

\begin{itemize}
	\item We consider all possible intersections $x \uplus Y$ and handle them sequentially.
	\item For each $y \in (x \uplus Y)$, we create an element $z$. The space covered by $z$ is the intersection of the spaces covered by $x$ and $y$, and the value is
$V(x \otimes y)$ (Definition~\ref{DefMetric}).
	\item There will be a ``remainder" when the process is finished if $x$ only has a partial intersection with $Y$. Thus, we remove from $x$ the space of each $z$ resulting from the intersection, and we add the remainder, if any, to the result.
\end{itemize}

After applying all three merge principles for each element $x \in X$, we process the elements $y \in Y$. Merge principles (1) and (2) have to be applied to these elements, but principle (3) can be partially avoided. Indeed, the elements in the intersection of $X$ and $Y$ have already been computed when processing $X$. We use a hash map $H$ as a cache, in which the element $y$ is used as a key and its corresponding value $y'$ in the hash map is the space that is updated throughout the process. When an intersection between $x$ and $y$ is found, the part in the intersection is virtually removed from $y$ by changing the value for $y$ in $H$. After all of the intersections have been computed, $H$ contains the remainder of each element of $Y$ and thus it can be added directly to the result.
	
{\bf Example.} In Figure~\ref{4-example}, the left decision tree was trained on a dataset with the degree ranging from 0 to 15, and the age ranging from 0 to 8. The right decision tree was trained on a dataset with the degree ranging from 0 to 11, and the age ranging from 0 to 9.
The light grey element in the left decision space and the element with the checkerboard pattern in the right decision space intersect. As two elements converted from decision trees only intersect in one contiguous space, the result is the new space $z$ whose age ranges from 7 to 8, and whose degree ranges from 3 to 10. Using Definition~\ref{DefMetric}, the percentage for the class \textit{Yes} is
$(\frac{15}{2} \times 40) / (\frac{15}{2} + \frac{13}{2}) + (\frac{13}{2} \times 0) / (\frac{15}{2} + \frac{13}{2}) \approx 21$,
and since there are only two classes in this example, the percentage for the class {\textsl No} is $100 - 21 = 79$.

\begin{figure}[t]
\centering
\includegraphics[scale=.35]{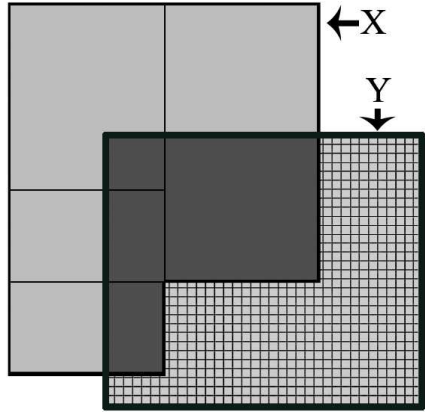}
\caption{Intersection of decision spaces $X$ (four rectangular elements) and $Y$ (one rectangular element).}
\label{4-FormeBizarre}
\end{figure}

The intersection of rectangles in {\em m}-dimensions (for {\em m} attributes) can result in spaces that are not rectangles. For example, Figure~\ref{4-FormeBizarre} shows the intersection of two decision spaces $X$ and $Y$.  Decision space $X$ contains four rectangular elements and $Y$ contains one rectangular element, but the intersection of the element of $Y$ with any element of $X$ leaves a non-rectangular remainder of $Y$. Constraining elements to be rectangles requires a heuristic that can bias the result since rectangles are only an approximation of the element's actual specialization.
We avoid this problem by using polytopes instead of rectangles to provide an exact algebra. The possibility that elements may have to be constrained to simple polytopes ({\em e.g.,} rectangles) for computational reasons is discussed in the Conclusions.

\subsection{Algebraic properties}
\label{algprop}

The goal of a merge operator $\otimes$ is to merge the information of two decision spaces, resolving any conflicts that arise. Thus, it has to obey a set of algebraic properties in order to be consistent:
\begin{itemize}
	\item {\em Commutativity:} merging a decision space with another one should not depend on which one is first but only on the information.
	\item {\em Identity element:} merging a decision space with a decision space that does not contain any elements should not change anything because there is no new information.
	\item {\em Idempotence:} merging a decision space with itself should not change anything, as there are neither conflicts nor new information.
\end{itemize}
Theorems~\ref{Th2},~\ref{Th3}, and~\ref{Th4} prove that our merge operator satisfies these three algebraic properties. Intuitively, the algorithmic construction of the merge operator is based on unions of geometric spaces and the resolution of conflicts encountered for non-empty intersections. Both the union of spaces ({\em i.e.,} the intersection of geometrical elements) and the resolution of conflicts ({\em i.e.,} the weighted average of values) satisfy the three algebraic properties. Therefore, we show that they are combined ``appropriately'' by the merge operator. We then show in Theorem~\ref{Th5} that the operator is not associative, which is the subject of the next section.

\begin{defn}
{\rm The set of all decision spaces is denoted by $\mathbb{D}$.}
\label{Def6}
\end{defn}

\begin{thm}
The $\otimes$ operator is commutative, {\em i.e.,} $\forall X,Y \in \mathbb{D}$, $X \otimes Y = Y \otimes X$.
\label{Th2}
\end{thm}

\begin{proof}Suppose that there is a $z \in X \otimes Y$. We prove that it also has to be in $Y \otimes X$. We consider all possible cases from which such a $z$ could result:
\begin{itemize}
	\item[(1)] $z$ results from an $x \in X$ that has no intersection. We show that if a $y \in Y$ has no intersection it will also be kept in the result $Z$.
	\item[(2)] $z$ results from the intersection of an $x \in X$ with a $y \in Y$. We will show that there are no changes if we consider it as the intersection of a $y \in Y$ with an $x \in X$.
	\item[(3)] $z$ is the remainder of an element $x \in X$. We show that the remainder of an element $y \in Y$ will also be kept in the result $Z$.
\end{itemize}
(1) As $x \in X$ has no intersection, it will be added to the result (line 6). If a $y \in Y$ has no intersection, it will not be in any $x \uplus Y$ and thus we will not create a $(y, y') \in H$ in line 15. As $(y, y') \notin H$ in line 22, $y$ will be considered unchanged. Given that $y$ has no intersection, it will be added to the result in line 24.\\
(2) As $x$ intersects with a $y$, the operator $\uplus$ relates $x$ to $y$ (line 9) and an element $z$ will be created, its space being the intersection of the spaces of $x$ and $y$, and its value being $V(x \otimes y)$. Each operation involved is commutative, thus the overall process is commutative.\\
(3) For an $x \in X$, we consider each $y \in Y$ with which it intersects (line 9): an element $z$ is created for each intersection, and its space is taken out of the space of $x$ (line 19); once the spaces of all intersections have been taken out of $x$, the remainder is added to the result if it is not empty (line 21). The same process takes place for the remainder of a $y \in Y$: all the $x \in X$ with which it intersects are considered (lines 3 and 9), an element $z$ is created for each intersection, and we keep track of the remainder of $y$ by updating its associated value in $H$ or creating it for the first intersection (lines 14--18). An element $y$ will not be considered anymore if it intersected with some $x \in X$ (line 25), and instead its remainder is added to the result (line 27).\label{Proof2}\end{proof}

\begin{defn}
{\rm A decision space $E \in \mathbb{D}$ is called an {\em identity} of $\mathbb{D}$ with respect to the $\otimes$ operator if and only if $\forall D \in \mathbb{D}, E \otimes D = D \otimes E = D$.}
\label{Def7}
\end{defn}

\begin{thm}
There exists a unique identity $E \in \mathbb{D}$ with respect to the $\otimes$ operator, and it is an empty set of polytopes ({\em i.e.,} the empty space).
\label{Th3}
\end{thm}

\begin{proof}Let $X$ and $E$ be two decision spaces and $E = \emptyset$. We first prove that $\forall X \in \mathbb{D}, X \otimes E = X$, which leads to $E \otimes X = X$ using Theorem~\ref{Th2}, hence E is an identity element. No element $x$ is subsumed by an element $e \in E$, and $x \uplus E = \emptyset$, thus all elements $x \in X$ are added to the result (line 6). As there is no $e \in E$, the loops in lines 22--27 are not executed and so the final result is equal to $X$.

We complete the proof by showing that $Y \in \mathbb{D}$ cannot be an identity for any $X \in \mathbb{D}$ if $Y \neq E$. As $Y \neq E = \emptyset$, there is at least one $y \in Y$. Let us consider $X \in \mathbb{D}$ such that $y \uplus X = \emptyset$. As $y$ has no intersection with any $x \in X$, it will be ignored by the main for loop (lines 3--21), thus $\nexists (y, y') \in H$. As a consequence, it will be considered by the second loop (line 22) and, as it is not subsumed by any $x \in X$, it will be added to the result in line 24; thus, the result is $X \cup y \neq X$ because $y \notin X$. Similarly, for the case $y \uplus X \neq \emptyset$, the element $y$ is processed by the main loop (lines 8--21). For any such $y$, $\exists X \in \mathbb{D}$ with an element $x \in X$ intersecting with $y$ such that $x$ and $y$ cover a different space. This results in dividing $x$ into new elements (line 11) and thus the result of $X \otimes Y$ is different from $X$.\label{Proof3}\end{proof}

\begin{thm}
The $\otimes$ operator is idempotent as $X \otimes X = X$.
\label{Th4}
\end{thm}

\begin{figure}[t]
\centering
\includegraphics[scale=.35]{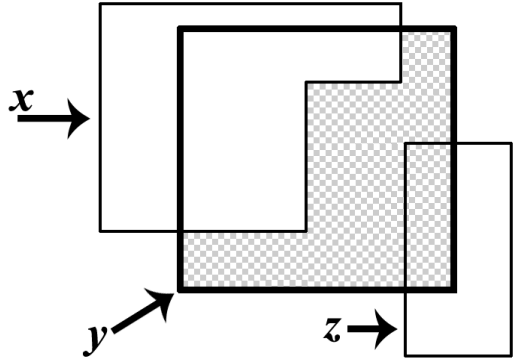}
\caption{Illustration of Theorem~\ref{Th5}.}
\label{4-associativity}
\end{figure}

\begin{proof}
\label{Proof4}
Let $X$ and $Y$ be two decision spaces such that $X = Y$:
\begin{itemize}
	\item There is no $x \in X$ strictly contained in an $y \in Y$. The use of strict subsumption $\prec$ instead of $\preceq$, in the definition of $\otimes$, is particularly important at this point. Indeed, if $X = Y$ and we were using $\preceq$, then all elements in $X$ would be discarded because they are subsumed by the same value in $Y$, and similarly for $Y$. Thus, $X \otimes X$ would lead to the incorrect result $\emptyset$.
	\item Each element $x$ intersects with exactly one $y$. As they are the same, we will add one element to the result with the following value for each component of the vector (lines 8--13):
\begin{center}$V(x) \times \frac{M(x)}{M(x)+M(y)} + V(y) \times \frac{M(y)}{M(x)+M(y)} = V(x) \times \frac{M(x)}{2 \times M(x)} + V(x) \times \frac{M(x)}{2 \times M(x)} = \mbox{V(x).} $\end{center}
	\item $x$ will be added to the result for all $x \in X$ (line 13). For all $y \in Y$, we added a pair $(y, z) \in H$ such that $z.space = x.space \setminus y.space = \emptyset$ (line 15). Thus, all elements of $y \in Y$ are skipped because they are in $H$ (line 22), and because the associated value is empty (line 26). Therefore the resulting decision space contains each $x \in X$ exactly once.
\end{itemize}
\vspace*{-0.7cm}
\end{proof}

\begin{cor}
{\rm The binary operator $\otimes: (\mathbb{D}, \mathbb{D}) \mapsto \mathbb{D}$ is idempotent, commutative, and contains a unique identity element for the set $\mathbb{D}$ of all decision spaces. Therefore, $(\mathbb{D},\otimes)$ is a unital, idempotent, and commutative magma (see~\cite{MS01} for a brief review of algebraic structures such as magmas).}
\label{Cor}
\end{cor}

\begin{thm}
The $\otimes$ operator is not associative: 
$\exists X,Y,Z \in \mathbb{D}$ such that
$(X \otimes Y) \otimes Z \neq X \otimes (Y \otimes Z)$.
\label{Th5}
\end{thm}

\begin{proof}
In Figure~\ref{4-associativity}, if we first merge $x$ with $y$, then the value of the intersection will depend on all of $x$ and $y$, and we get the shaded remainder of $y$. Then, if we merge with $z$, the value of the resulting intersection will depend on the shaded remainder and $z$. However, if we first merge $y$ with $z$, then the part of $y$ that intersects with $z$ is removed from $y$, resulting in a remainder $y'$. Then, if we merge with $x$, the value depends on $x$ and $y'$ which covers a smaller space than $y$. It follows that its intersection with $x$ will have a different value. Thus, the values change with the order in which elements are merged, while the spaces covered do not (as they result from the intersection of geometrical spaces which is an associative operation).
\label{Proof5}\end{proof}

\section{The impact of merge order}
\label{time}

According to Theorem~\ref{Th5}, the $\otimes$ operator is not associative, so the result of merging more than two decision spaces depends on the order in which they are merged. We refer to a merge order as a {\em merging scheme}. Note that a scheme is not necessarily a simple sequence: a merging scheme can specify that large groupings of decision spaces have to be merged first, and then merged together. Its representation is defined below.

\begin{defn}
A {\em merging scheme} specifies the order in which a set of decision spaces is merged. It can be represented as a tree, in which a leaf represents a decision space, an intermediate node represents the application of the $\otimes$ operator (hence an intermediate decision space), and the root represents the final result.
\end{defn}

\begin{figure}
\centering
\includegraphics[scale=.35]{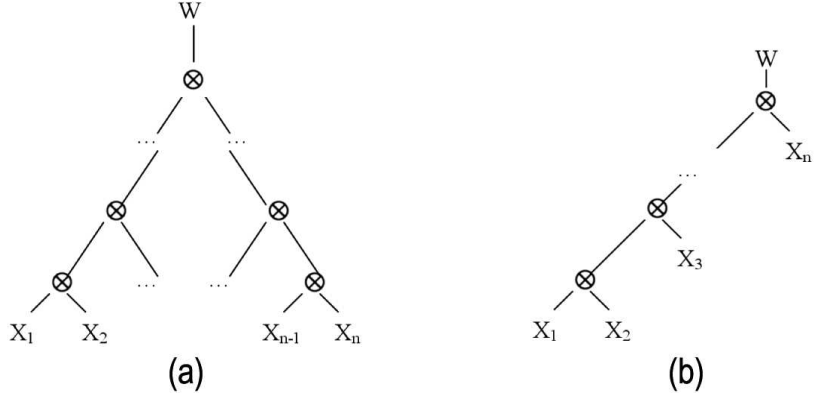}
\caption{Unbiased (a) and biased (b) binary merging schemes.}
\label{4-unfair}
\end{figure}

Two merging schemes are shown in Figure~\ref{4-unfair}. In the merging scheme in~\ref{4-unfair}(a), the decision spaces are merged pairwise (bottom layer of the tree), then the resulting decision spaces are merged pairwise, and so on, until a single decision space $W$ is obtained. We can describe this merging scheme as $( ( (X_1 \otimes X_2) \otimes (X_3 \otimes X_4) ) \cdots ( (X_{m-3} \otimes X_{m-2}) \otimes (X_{m-1} \otimes {X_m}) ) )$, where $m = 2^k$. As we will demonstrate in Corollary~\ref{Cor6}, this merge order does not introduce any bias into the calculation of the value of $W$. Any differences in the impacts of the decision spaces on the value of $W$ are based on their relative specializations. In contrast, the merging scheme in Figure~\ref{4-unfair}(b) will lead to a bias in the result. In this scheme, the first two decision spaces are merged, then the result is merged with the third decision space, and so on, until $W$ is obtained. We can describe this merging scheme as $(((X_1 \otimes X_2) \otimes X_3) \cdots \otimes X_m)$. This merge order does introduce bias into the calculation of the value of $W$. Each decision space is merged with the result of merging its predecessors, it's impact is the same as all of its predecessors combined. So, the impact of earlier decision spaces is reduced each time that a new decision space arrives. This effect can be desirable. If the underlying distribution is changing, then we often want recent decision spaces to have a larger impact on the result because they represent recent trends.

Both merging schemes can be desirable depending on the setting. The merging scheme shown in Figure~\ref{4-unfair}(a) can be used when a homogeneous database has been partitioned among several computation units. Indeed, the model produced by each unit should have the same impact on the final result as the underlying distribution is homogeneous. The merging scheme shown in Figure~\ref{4-unfair}(b) can be used for time-changing distributions, for example data streams. In this setting, learners can deduce models at different times, and we may want to favour the most recent models as they represent recent trends. The situation in which decision spaces are merged as soon as they are received is depicted in Figure~\ref{4-unfair}(b): the decision space received at time $t$ is merged with everything that was received up to time $t-1$. Note that we could also have an online setting such as in data streams, but we might not want the time at which models are received to have an impact. In this case, the scheme in Figure~\ref{4-unfair}(a) can also be used, and it specifies that some decision spaces will have to be stored temporarily: large groupings of decision spaces must be formed before the merging operation can take place to guarantee an equal impact on the final result.

However, the framework developed so far has limitations that can prevent unbiased results. Indeed, the merging operator is binary and thus an unbiased result can only be achieved if the number of decision spaces to merge is a power of 2. Furthermore, using powers of 2, the impact of decision spaces over time can only decay exponentially, whereas other types of decays may be needed. We will generalize the merge operator to overcome these limitations in the next two subsections.

\subsection{$m$-ary merging}

Consider a set $\{X_1, \ldots, X_m\}$ of decision spaces. 
We extend Algorithm~2 
 from a binary operator to an $m$-ary operator. Increasing the number of geometrical objects that are intersected does not affect the operator, since the intersection of geometrical objects is associative. Thus, the extension only involves the computation of the value, which is extended in Definition~\ref{DefMultiMetric}.
\begin{defn}
The {\em value} of the intersection of $m$ elements $x_1 \in X_1, x_2 \in X_2, \ldots,$ $x_m \in X_m$ based on their specializations is a vector
\[V(x_1 \otimes \cdots \otimes x_m) = V(x_1) \times \frac{M(x_1)}{M(x_1)+ \cdots +M(x_m)} + \cdots + V(x_m) \times \frac{M(x_m)}{M(x_1)+ \cdots +M(x_m)}.\]
\label{DefMultiMetric}
\end{defn}

The combination of the handling of geometrical spaces and the computation of the value is similar to the merging algorithm introduced in Section~\ref{merge2}:
\begin{itemize}
	\item[(1)] If an element is strictly included within another one and has the same value, discard it.
	\item[(2)] Otherwise, for each intersection, create a new element $z$ with a value that is computed using the formula in Definition~\ref{DefMultiMetric}.
\end{itemize}

We assume in Theorem~\ref{Th8} and its corollaries that the decision spaces to be merged all cover the same space so that we can concentrate on the impact of merge order.

\begin{thm}
In a merging scheme for $m$ decision spaces $X_1, \ldots, X_m$, all of which cover the same space, the impact of a decision space $X_i$ on the value of the final result is proportional to the product of the numbers of operands of the $\otimes$ operators on the path from the leaf representing $X_i$ to the root of the merging scheme.
\label{Th8}
\end{thm}

\begin{proof}
In a merging scheme, each internal node is an $m$-ary $\otimes$ operator and each leaf is a decision space.  Consider an internal node $u$ with $k$ operands $T_1, T_2, \ldots, T_k$. Each $T_j$ is either a decision space or a subtree of the merging scheme and each $T_j$ accounts for a fraction $1/k$ of the value of the subtree rooted at $u$.  Consider two internal nodes $u_1$ and $u_2$ with $k_1$ and $k_2$ operands, respectively, and suppose that $u_2$ is an operand of $u_1$ in the merging scheme.  Then each operand of $u_2$ accounts for $1/k_2$ of its value and $u_2$ accounts for $1/k_1$ of the value of $u_1$, so each operand of $u_2$ accounts for a fraction $1/(k_1 \times k_2)$ of the value of $u_1$.  If the numbers of operands of $u_1$ and $u_2$ are exchanged, then each operand of $u_2$ contributes the same fraction $1/(k_2 \times k_1)$ of the value of $u_1$. Thus, the contribution of a decision space to the value of the root of the merging scheme depends only on the final product of the numbers of operands of the $m$-ary $\otimes$ operators on the leaf to root path.
\label{Proof7}\end{proof}

\begin{figure}
\centering
\includegraphics[scale=.5]{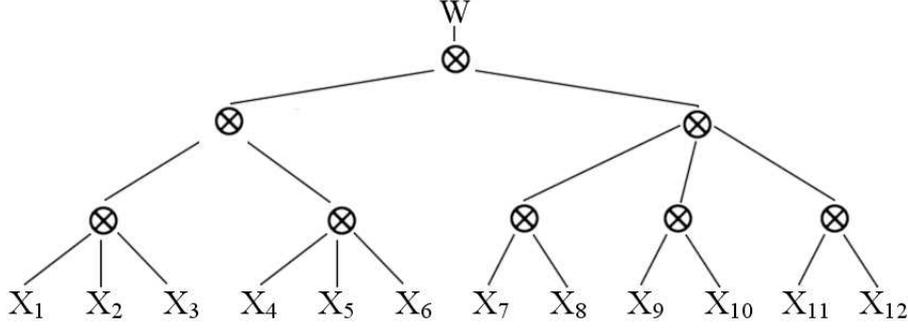}
\caption{Merging scheme from Theorem~\ref{Th8} with $n$ = $3 \times 2 \times 2$ decision spaces.}
\label{4-optimal}
\end{figure}

Figure~\ref{4-optimal} shows an example of a merging scheme for twelve decision spaces in which the impacts of all of the decision spaces on the value of $W$ are the same.

\begin{cor}
In the merging scheme $( ( (X_1 \otimes X_2) \otimes (X_3 \otimes X_4) ) \cdots ( (X_{n-3} \otimes X_{n-2}) \otimes (X_{n-1} \otimes {X_m}) ) )$, where $m = 2^k$ (Figure~\ref{4-unfair}(a)), each decision space accounts for $2^{-k}$ of the value of $W$.
\label{Cor6}
\end{cor}

\begin{cor}
In the merging scheme $(((X_1 \otimes X_2) \otimes X_3) \cdots \otimes X_m)$ (Figure~\ref{4-unfair}(b)), decision space $X_i$ accounts for $2^{-m+i-1}$ of the value of $W$ if $i > 1$, and $2^{-m+1}$ if $i = 1$.
\label{Cor7}
\end{cor}

\subsection{Adjusting the bias}

As explained earlier in this section, when decision spaces are merged over time as in a data stream setting, then their impacts decay exponentially. In the previous sub-section, we used extra storage to change the impacts. In this sub-section, we introduce an approach that uses no extra storage and provides decision spaces with the same impact. Furthermore, this approach does not require prior knowledge of the number $m$ of decision spaces to be merged. In contrast, the approach in the previous sub-section requires $m$ to design a merging scheme.

\begin{defn}
The {\em $\beta$-weighted value} of the intersection of two elements $x \in X$ and $y \in Y$ is the class distribution vector
$V(x \otimes y) = V(x) \times \beta + V(y) \times (1-\beta)$, $0<\beta<1.$
\label{DefWMetric}
\end{defn}

The value produced by the formula in Definition~\ref{DefMetric} is a $\beta$-weighted value with $\beta=\frac{M(x_1)}{M(x_1)+M(x_2)}$ and $V(x \otimes y)$ depends only on the specializations of $x$ and $y$. If $x$ and $y$ have the same specialization, then each accounts for half of the value of the result. The bias in a merging scheme like the one in Figure~\ref{4-unfair}(b) is entirely the result of the merge order when $\beta=\frac{M(x_1)}{M(x_1)+M(x_2)}$ is used. We can use different values of $\beta$ with this scheme to produce the final value $V(((x_1 \otimes x_2) \otimes x_3) \cdots \otimes x_m) = V(x_1) \times \frac{M(x_1)}{M(x_1)+ \cdots +M(x_m)} + \cdots + V(x_m) \times \frac{M(x_m)}{M(x_1)+ \cdots +M(x_m)}$, the same as an $m$-ary merge.

\begin{thm}
Let $\beta_i=\frac{M(x_1)+\cdots +M(x_i)}{M(x_1)+\cdots +M(x_{i+1})}$ be the weight used to compute the value of the $i$-th merge, $i=1,2,\ldots,m-1$, in the merging scheme $(((X_1 \otimes X_2) \otimes X_3) \cdots \otimes X_m)$. Then the resulting value of the intersection of $m$ elements $x_1 \in X_1, x_2 \in X_2, \ldots, x_m \in X_m$ based on their specializations is $V(((x_1 \otimes x_2) \otimes x_3) \cdots \otimes x_m) = V(x_1) \times \frac{M(x_1)}{M(x_1)+ \cdots +M(x_m)} + \cdots + V(x_m) \times \frac{M(x_m)}{M(x_1)+ \cdots +M(x_m)}$.
\end{thm}

\begin{proof}
We prove this result by induction on $m$. If $m=1$, then two decision spaces are merged using $\beta_1=\frac{M(x_1)}{M(x_1)+M(x_2)}$ and $V(x_1 \otimes x_2) = V(x_1) \times \frac{M(x_1)}{M(x_1)+M(x_2)} + V(x_2) \times (1-\frac{M(x_1)}{M(x_1)+M(x_2)}) = V(x_1) \times \frac{M(x_1)}{M(x_1)+M(x_2)} + V(x_2) \times \frac{M(x_2)}{M(x_1)+M(x_2)}$. Assume that the theorem is true for $m=j-1\geq 2$, so that $V((((x_1 \otimes x_2) \otimes x_3) \cdots \otimes x_j) = V(x_1) \times \frac{M(x_1)}{M(x_1)+ \cdots +M(x_j)} + \cdots + V(x_j) \times \frac{M(x_j)}{M(x_1)+ \cdots +M(x_j)}$. The $j$-th merge uses $\beta_j=\frac{M(x_1)+\cdots +M(x_j)}{M(x_1)+\cdots +M(x_{j+1})}$. Then $V((x_1 \otimes \cdots \otimes x_j) \otimes x_{j+1}) = (V(x_1) \times \frac{M(x_1)}{M((((x_1 \otimes x_2) \otimes x_3) \cdots \otimes x_j)) \otimes x_{j+1}} + \cdots + V(x_j) \times \frac{M(x_j)}{M(x_1)+ \cdots +M(x_j)}) \times \frac{M(x_1)+\cdots +M(x_j)}{M(x_1)+\cdots +M(x_{j+1})} + V(x_{j+1}) \times (1 - \frac{M(x_1)+\cdots +M(x_j)}{M(x_1)+\cdots +M(x_{j+1})}) = V(x_1) \times \frac{M(x_1)}{M(x_1)+ \cdots +M(x_{j+1})} + \cdots + V(x_{j+1}) \times \frac{M(x_{j+1})}{M(x_1)+ \cdots +M(x_{j+1})}$.
\end{proof}

\section{Developing an algebra}
\label{algebra}

\subsection{Restriction operator}

In a time-varying environment, the accuracy of the predictions decreases over time, so new decision spaces must be created regularly. A sequence of decision spaces carries information about changes in the environment over time, so techniques such as time series analysis potentially could be used. While time series analysis considers a sequence of vectors of {\em fixed size}, decision spaces can be of varying size: for example the values of attributes can evolve over time to cover a broader space. A restriction operator can be used to simplify the decision spaces of a time series so that they all have same size. We introduce this operator in Definition~\ref{defrestriction}, and define it formally in Algorithm~3. 
Then, we apply the algebraic approach in a similar way to the merging operator: we study identity elements (Theorem~\ref{Th10}), idempotency (Theorem~\ref{Th11}), and associativity (Theorem~\ref{Th12}).

\begin{defn}
The {\em restriction} of a decision space $X$ by a decision space $Y$ is the decision space $Z$ that only retains elements of $X$ that intersect with elements of $Y$. The corresponding operator is denoted $\odot$.
\label{defrestriction}
\end{defn}

\noindent
\begin{minipage}{\textwidth}
\label{Alg3}
\newlength{\algcomc}
\setlength{\algcomc}{8cm}
\hrulefill\\
{\bf Algorithm 3}\quad $\odot: (X, Y) \mapsto Z$\\

\vspace*{-6mm}\hrulefill\\
\vspace*{-5mm}\begin{tabbing}
xxx\=xx\=xx\=XXXXXXXXXXXXXXXXXXXXXXXXX\= \kill

1.\>$Z \leftarrow$ new decision space\\
2.\>{\bf for} $x \in X$ such that $x \uplus Y \neq \emptyset$ {\bf do}\\
3.\>\>$z \leftarrow$ new element\\
4.\>\>{\bf for} $y \in x \uplus Y$ {\bf do}\\
5.\>\>\>$z.space \leftarrow z.space \cup (x.space \cap y.space)$\\
6.\>\>$z.value \leftarrow x.value$\\
7.\>\>$Z \leftarrow Z \cup z$\\
8.\>{\bf return}$Z$\\
\end{tabbing}
\end{minipage}

By definition, we only consider the elements $x \in X$ that intersect with some elements $y \in Y$ (line 2). For each such element $x$, we create an element $z$ with the same value (line 6) but with a space restricted to the intersection between $x$ and all $y \in Y$ (lines 4--5). Intuitively, restriction is the intersection of geometrical spaces, and it leaves the values unchanged. Thus, the algebraic properties of restriction are derived only from the properties of intersection of spaces.

\begin{thm}
Let $F_v$ be a family of decision spaces such that for every $X \in F_v$, there is exactly one element $x \in X$ such that $V(x) = v$ and $\forall a \in A(x), S(x,a) = \infty$. Then, every identity element $E \in \mathbb{D}$ such that $X \odot E = E \odot X = X$ is generated by $F_v$. The number of elements of $F_v$ is infinite.
\label{Th10}
\end{thm}

\begin{proof}A decision space $X$ is not restricted by a decision space $Y$ only if all elements in $Y$ cover a space at least as large as the space covered by $X$, so a trivial identity is the decision space with only one element covering an infinite space. As the value of this element does not matter, we can define a \textit{family} taking its value $v$ as a parameter. The set of all possible values $v$ can be infinite because the values are taken from a continuous range; thus, $F_v$ contains an unbounded number of identity decision spaces.\label{Proof9}\end{proof}

\begin{thm}
The $\odot$ operator is idempotent as $X \odot X = X$.
\label{Th11}
\end{thm}

\begin{proof}Let $X$ and $Y$ be two decision spaces. If $X = Y$ then $\forall x \in X$, $x \uplus Y = y$ such that $\forall a \in A(x)$, $low(a, x) = low(a, y)$ and $up(a, x) = up(a, y)$. Thus, each element $z$ is created with exactly the space and value of an $x$ (lines 4--7), and only one $z$ is created for each $x$ (line 2). Hence the result $Z$ is the same as $X$.\label{Proof10}\end{proof}

\begin{thm}
The operator $\odot: (\mathbb{D}, \mathbb{D}) \mapsto \mathbb{D}$ is associative:
$(X \odot Y) \odot Z = X \odot (Y \odot Z)$ for all decision spaces $X, Y, Z \in \mathbb{D}$.
\label{Th12}
\end{thm}

\begin{proof}As the values do not matter, the restriction can be considered to be an intersection of spaces, which is associative.\label{Proof11}\end{proof}

\subsection{Composite operators}

The merge ($\otimes$) and restriction ($\odot$) operators can be composed to create a variety of useful operators. When the merge operator creates an element $z \in Z = X \otimes Y$, $z$ can be created from an $x \in X$, a $y \in Y$, or both an $x \in X$ and a $y \in Y$. The $\odot$ operator can be used to restrict the results of a merge operation to include only elements $z \in Z$ that are created from both an $x \in X$ and a $y \in Y$. This change has a significant impact: an element derived from only one element is not as precise as an element derived from two, because in the latter case a consensus is obtained through a weighted formula. Thus, this restricted form of merging is less sensitive to noise and, as we know that all elements in $Z$ are derived from exactly two elements, we can have the same confidence in the predictions of all elements in $Z$. A merge that provides the same confidence in the prediction of each element is obtained through the following composite operator $\oplus$.

\begin{defn}
The operator $\oplus:(X,Y) \mapsto Z$ is defined by $X \oplus Y = (X \otimes Y) \odot X \odot Y$.
\label{Def11}
\end{defn}

This definition ensures that the values are correctly computed based on the specialization of each element of $X$ and $Y$, and then the overall space is reduced to the intersections with $X$ and $Y$. A stricter definition is given by the following composite operator $\bar{\odot}$ that not only restricts the merging to the elements that intersect, but computes the values using only common spaces of attributes. Any part of an element that lies outside the intersection will be ignored by $\bar{\odot}$ when measuring the specialization, and the values are computed based on the same space.

\begin{defn}
The operator $\bar{\odot}: (X, Y) \mapsto Z$ is defined by $X \bar{\odot} Y = (X \odot Y) \otimes (Y \odot X)$.
\label{Def12}
\end{defn}

The choice between $\bar{\odot}$ and $\oplus$ can be based on the application. Both $\bar{\odot}$ and $\oplus$ have the properties of idempotency, associativity, non-commutativity, and unique identity element.

\section{Conclusions}
\label{conclusions}

Classifiers are found in a variety of settings, such as sensor networks. In these settings, each entity ({\em e.g.,} a sensor) deduces a classifier from its observations (i.e., a dataset of example instances). These classifiers must be merged to obtain a global view. However, merging has not been defined for several types of classifiers, such as decision trees. Furthermore, the classifiers that have to be merged can be of different types, for example, when several generations of sensors are creating classifiers.

In this paper, we introduced \textit{decision spaces} as a framework for examining the properties of operations over classifiers, and we defined a merge operation that handles possibly different classifiers. Our methods are designed for a pure meta-learning framework in which classifiers are merged instead of being stored separately, and observations are \textit{never} exchanged. We use algebraic methods to prove several desirable properties of the merge operator. We also show that the operator is not associative: the result of merging several classifiers depends on the order that the merge operator is applied. This can be the desired behaviour in a situation such as a data-stream: if the operator is applied as soon as a classifier is received, then the result will be biased in favour of recent trends. However, other types of biases may be needed, and no bias should also be possible for an application such as a distributed database in which the underlying data is homogeneous. We introduced two approaches to achieve different biases. The first one uses storage space flexibly to customize the bias. The second approach results in no bias, thereby making the operator associative, and uses no space.

We showed that other operators can be defined in our framework, and used algebraic methods to establish their behaviours. In particular, we defined a restriction operator, which can be used for time series to analyze a sequence of classifiers, and can be used in combination with the merge operator for application-specific needs.

The decision space framework relies on the intersection of geometrical elements. However, intricate shapes can result from the intersections of large numbers of attributes and/or classifiers. This can affect the space complexity since the shapes may require more coordinates to be described, and it can affect the time complexity as the shapes become harder to intersect. One direction for further research is the use of simplified shapes to reduce the time and space complexities. This suggests several interesting questions: how do the time and space complexities depend on the type of geometrical elements that are allowed? How does the accuracy of the results depend on the simplification that is chosen and when it is applied?

Further questions arise when considering that uncertainty is found at many levels in real-world scenarios. We distinguish three broad types of uncertainty. Firstly, there can be uncertainty about the concept that the elements are predicting. For example, consider the task of predicting the literary genre of an author based on dimensions measuring cultural features. Since the definition of a genre may not be clear, fuzzy values may have to be assigned to elements, and the problem becomes one of combining fuzzy values.

\begin{figure}
\centering
\includegraphics[scale=.33]{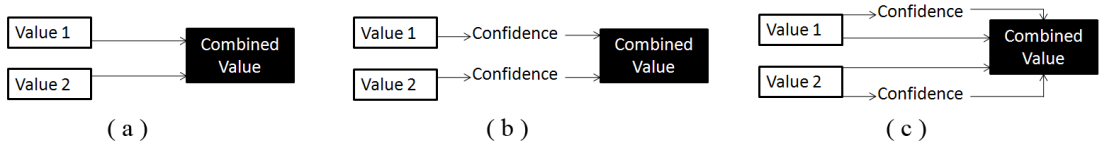}
\caption{Including confidence in a combined value.}
\label{Fig:Confidence}
\end{figure}

Secondly, there can be uncertainty about the value contained in one element (i.e., its class distribution vector). Consider the case in which two elements $A$ and $B$ are merged: element $A$ predicts classes Yes and No with the same probability, and element $B$ predicts the class Yes with 80\% probability and No with 20\% probability. A mechanism is needed to detect that element $A$ cannot make a reliable prediction, so the contribution of element $B$ should be increased. Confidence transformations provide such a mechanism by transforming a class distribution vector into a measure representing its confidence~\cite{LHS04}. There are several approaches to using these transformations when combining two elements in this framework. In Figure~\ref{Fig:Confidence}(a), two values (i.e., class distribution vectors) are combined based solely on the space of the elements to which they belong, ignoring their confidence measures. In Figure~\ref{Fig:Confidence}(b), the values are transformed to confidence measures which are then used to generate the combined value. Another possibility is to use the confidence measures to supplement the raw values (Figure~\ref{Fig:Confidence}(c)).

A confidence transformation can be specific to a given classifier (i.e., we interpret the reliability of a value differently depending on the type of model that generated it). However, that specificity is lost when we combine a set of classifiers: how should we interpret the result coming from a classifier that combines a C4.5 decision tree with a support vector machine? If one general confidence transformation is used for such models, then how does the performance change when a large number of models are combined based on initially large groupings (so that specific transformations can be used)? Is it possible to not only combine models, but also combine the transformations associated with them?

Thirdly, there can be uncertainty about the space covered by an element. For example, consider the classification of soil types on a map. In one area, samples have been taken and the category has been determined. It is known that beyond that area, there are no geological faults and it can thus be speculated that the category does not change for a few dozen meters beyond the sampled area. This type of spatial vagueness is essential in geographical information systems, and frameworks have been proposed to address it. We can adopt the idea of an algebra for vague spatial data from~\cite{PS10} (which further highlights the connection of our approach with spatial algebras), by tagging each element of a decision space as being either guaranteed or conjectured. The tag for elements resulting from the union, intersection, or difference of two elements can be determined by using the same tables as in~\cite{PS10}. However, what should we do when merging an uncertain element with a guaranteed element? Should we omit the value from the uncertain element, or should we produce a weighted combination giving more weight to the value from the guaranteed element? The answers to these questions are application-specific, and the important advantage of our framework lies in its ability to mathematically express the consequences of each choice.

\section*{Acknowledgements}
\label{thanks}

We would like to thank Martin Ester, Binay Bhattacharya, and Oliver Schulte for helpful discussions. This research was supported by NSERC of Canada.


\bibliographystyle{model1b-num-names}
\bibliography{giab-pet}

\end{document}